# Commutativity and Commutative Pairs of Some Differential Equations

**Mehmet Emir Koksal**

Department of Mathematics, Ondokuz Mayis University, 55139 Atakum, Samsun, Turkey

emir_koksal@hotmail.com

**ABSTRACT**

In this study, explicit differential equations representing commutative pairs of some well-known second-order linear time-varying systems have been derived. The commutativity of these systems are investigated by considering 30 second-order linear differential equations with variable coefficients. It is shown that the system modeled by each one of these equations has a commutative pair with (or without) some conditions or not. There appear special cases such that both, only one or neither of the original system and its commutative pair has explicit analytic solution. Some benefits of commutativity have already been mentioned in the literature but a new application for in cryptology for obscuring transmitted signals in telecommunication is illustrated in this paper.

# 1. INTRODUCTION

As one of the main fields of applied mathematics, differential equations arise in acoustics, electromagnetic, electrodynamics, fluid dynamics, wave motion, wave distribution, and in many other sciences and branches of engineering. There is a tremendous amount of work on the theory and techniques for solving differential equations and on their applications [1-4]. Especially, they are used as a powerful tool for modelling, analyzing and solving real engineering problems and for discussing the results turned up at the end of analyzing for resolution of naturel problems. For example, they are used in system and control theory, which is an interdisciplinary branch of electric-electronics engineering and applied mathematics that deal with the behavior of dynamical systems with inputs, and how their behavior is modified by different combinations such as cascade and feedback connections [5-8]. When the cascade connection in system design is considered, the commutativity concept places an important role to improve different system performances [9-11].

When two subsystems are connected one after the other so that the output of the former excites the input of the later, this combination is known to be cascade connection. When the problem is which subsystem should appear first, this depends mainly on original properties of the specific system designed. However, when other criteria such as stability, sensitivity, noise disturbance, robustness, linearity effects are also considered, engineering ingenuity coupled with appropriate mathematical analysis play an important role on the decision. In any case, the interconnected subsystems as a whole must yield the same functioning that is, input-output relation, in ideal conditions. Therefore, the study of conditions under which the input-output relation of two time-varying subsystems connected in cascade is invariant of the order of connection (commutativity) is an important subject concerning engineers and mathematicians.

As shown in Figure 1, when the connection order of two cascade-connected continuous linear time-varying systems $A$ and $B$ is changed, if input-output relation of the assembled systems $AB$ and $BA$ does not change, we say that $A$ and $B$ are commutative systems and $(A, B)$ constitutes a commutative pair.

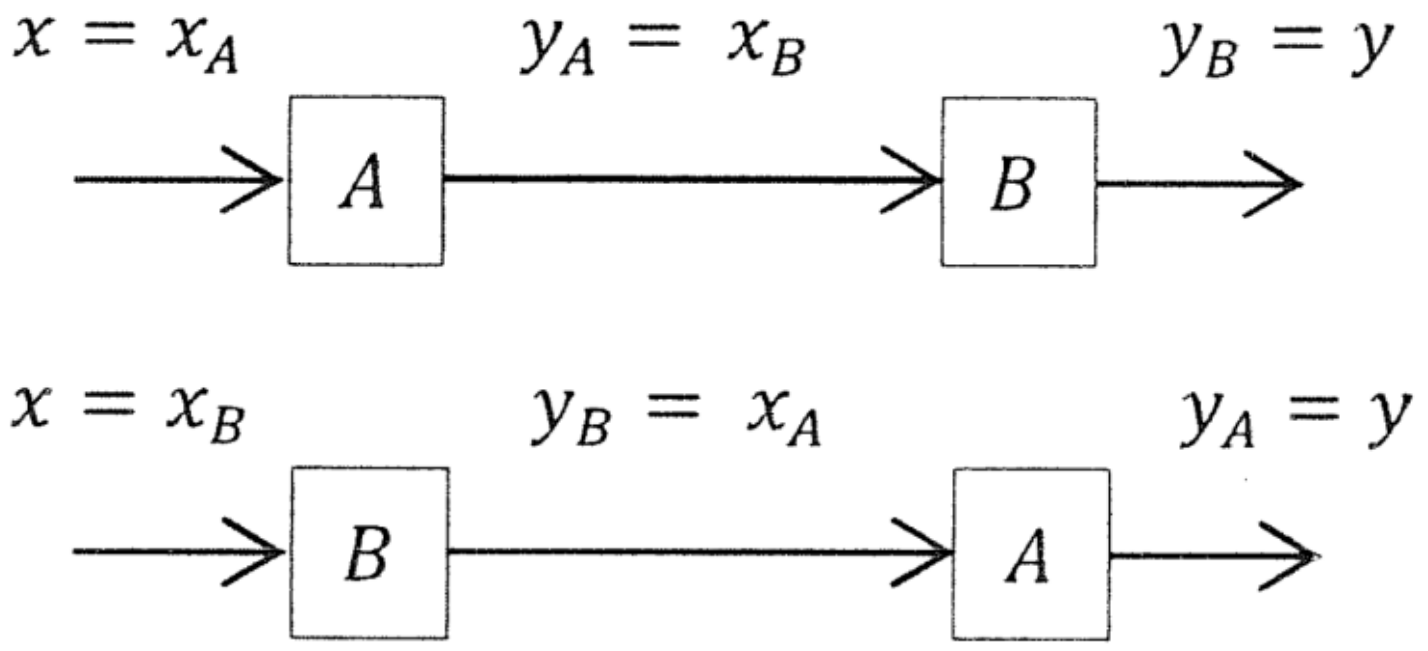

**Figure 1:** Cascade connection of the differential systems $A$ and $B$.

This subject was studied in the literature for the first time by E. Marshall in 1982 [1]. E. Marshall investigated commutativity of first-order differential linear time-varying system. He revealed very important reality that “a linear time-varying system can be commutative only with another linear time-varying system”. His study for first-order linear time-varying differential systems is simple but very important for moving this subject to the literature to reveal a new research area.

After that, commutativity has been investigated by a few scientists only and several important developments have been realized theoretically. M. Koksal [2-5] and S. V. Saleh [6] investigated commutativity of second-order linear time-varying systems and obtained necessary and sufficient conditions for their commutativity. Commutativity conditions of second-order linear time-varying systems were studied in [7] for more general cases. The previous results published in the literature and commutativity of fifth-order linear time-varying differential systems were presented and investigated in [8] by M. Koksal and M. E. Koksal.

Cascade-connected systems are important especially in electrical circuits [9, 10]. Moreover, linear time-varying circuits are among the fundamentals of modern communication used in modulation theory in electronics [11, 12]. So, commutativity of linear time-varying differential systems is also important in application.

Commutativity of cascade-connected linear time-varying systems was also studied in [13] for the first time for discrete time systems. Studies on commutativity of continuous-time (analog) linear time-varying systems light the way to the developments of the commutativity of discrete-time (digital) linear time-varying systems as future works.

In this study, many of the second-order linear differential equations with variable coefficient in the literature are reviewed in Chapter 2. In Chapter 3, the theoretical results of [7] are applied to these second-order differential equations for finding their commutative conjugates. In Chapter IV, the analytical solvability of the computed commutative conjugates are discussed. Chapter V includes a new application of commutativity for obscuring transmitted signal in a communication system. Finally, the paper ends up with Conclusions which constitute Section VI.

## 2. SECOND-ORDER DIFFERENTIAL EQUATIONS

We consider the linear time-varying system $A$ described by the following second-order linear time-varying differential equation

$$a_2(t)\ddot{y}_A(t) + a_1(t)\dot{y}_A(t) + a_0(t)y_A(t) = x_A(t); t \geq 0 \quad (1)$$

with initial conditions $y_A(0)$ and $\dot{y}_A(0)$. $a_2(t), a_1(t)$ and $a_0(t)$ are time-varying coefficients such that $a_2(t) \not\equiv 0$. $x_A(t)$ and $y_A(t)$ are the input and output functions of System $A$, respectively. Note that $\ddot{y}_A(t) = y_A''(t) = \frac{d^2}{dt^2}y_A(t)$, $\dot{y}_A(t) = y_A'(t) = \frac{d}{dt}y_A(t)$, and similar notations will be followed throughout the paper.

The following 30 second-order linear differential equations which are generally famed by the person who constructed these equations in the literature [14-17] are considered and listed in Table 1.

**Table 1:** Well-known second-order linear differential equations

| Line # | Name of Equation | Formula |
|---|---|---|
| 1 | Airy DE | $y'' \pm k^2xy = 0$ |
| 2 | Anger DE | $y'' + \frac{y'}{x} + \left(1 - \frac{v^2}{x^2}\right)y = \frac{x - v}{nx^2}sin(\pi v)$ |
| 3 | Baer DE | $(x - d_1)(x - d_2)y'' + 0.5[2x - (d_1 + d_2)]y' - (p^2x + q^2)y = 0$ |
| 4 | Bessel DE | $x^2y'' + xy' + (x^2 - n^2)y = 0$ |
| 5 | Bessel DE-modified | $x^2y'' + xy' - (x^2 + n^2)y = 0$ |
| 6 | Bessel DE-spherical | $x^2y'' + 2xy' + [x^2 - n(n + 1)]y = 0$ |
| 7 | Bessel DE-modified spherical | $x^2y'' + 2xy' - [x^2 + n(n + 1)]y$ |
| 8 | Bessel DE-wave | $x^2y'' + xy' + (a^2x^4 + b^2x^2 - c^2)y = 0$ |
| 9 | Chebyshev DE | $(1-x^2)y''-xy'+n^2y=0,$ where $\lvert x\rvert<1$ |
| 10 | Eckart DE | $y''+\left[\frac{\alpha\eta}{1+\eta}+\frac{\beta\eta}{(1+\eta)^2}+\gamma\right]y=0,$ where $\eta=e^{\delta x}$ |
| 11 | Ellipsoidal wave DE | $y''-(a+bk^2sn^2x+qk^4sn^4x)y=0,$ where $snx=sn(x,k)$ |
| 12 | Erfc DE | $y''+2xy'-2ny=0$ |
| 13 | Euler DE | $x^2y''+\alpha xy'+\beta y=s(x)$ |
| 14 | Gegenbauer DE | $(1-x^2)y''-2(\mu+1)xy'+(v-\mu)(v+\mu+1)y=0$ |
| 15 | Hill's DE | $y''+\left[\theta_0+2\sum_{n=1}^{\infty}\theta_n\cos(2nx)\right]y=0$ |
| 16 | Hypergeometric DE | $x(1-x)y''+[c-(\alpha+b+1)x]y'-aby=0$ |
| 17 | Jacobi DE-first | $(1-x^2)y''+[\beta-\alpha-(\alpha+\beta+2)x]y'+n(n+\alpha+\beta+1)y=0$ |
| 18 | Jacobi DE-second | $x(1-x)y''+[\beta-(\alpha+1)x]y'+n(n+\alpha)y=0$ |
| 19 | Laguerre DE | $xy''+(\alpha+1-x)y'+\lambda y=0$ |
| 20 | Magnetic Pole DE | $y''-\left[\frac{m(m+1)+0.25-(m+0.5)\cos x}{\sin^2 x}+\lambda+\frac{1}{2}\right]y=0$ |
| 21 | Morse-Rosen DE | $y''+\left[\frac{\alpha}{cosh^2(ax)}+\beta tanh(ax)+\gamma\right]y=0$ |
| 22 | Neumann DE | $x^2y''+3xy'+(x^2+1-n^2)y=x\cos^2\left(\frac{n\pi}{2}\right)+n\sin^2\left(\frac{n\pi}{2}\right)$ |
| 23 | Parabolic Cylinder DE | $y''+(\alpha x^2+bx+c)y=0$ |
| 24 | Riccati DE | $x^2y''+[x^2-n(n+1)]y=0$ |
| 25 | Richardson's DE | $-y''=(\lambda sgnx+\mu)y$ |
| 26 | Struve DE | $x^2y''+xy'+(x^2-v^2)y=\frac{4\left(\frac{x}{2}\right)^{v+1}}{\sqrt{\pi}\,\Gamma\left(v+\frac{1}{2}\right)}$ |
| 27 | Symmetric top DE | $y''-\left(\frac{M^2-0.25+K^2-2MK\cos x}{\sin^2 x}+\delta+K^2+\frac{1}{4}\right)y=0$ |
| 28 | Titchmarsh's DE | $y''+(\lambda-x^{2n})y=0$ |
| 29 | Weber DE-first | $y''+\left(-\frac{b^2}{4}x^2+a^2\right)y=0$ |
| 30 | Weber DE-second | $y'' + \frac{1}{x}y' + \left(1 - \frac{v^2}{x^2}\right)y = -\frac{1}{\pi x^2}[x + v + (x - v)cos(v\pi)]$ |

Some of the above equations are special or more general forms of the others; Moreover, some can be transformed to another. Which one is special or general form of another is not important in our study. We only consider the above famous linear differential equations for the existences of their commutative pairs; if they exist, finding commutative pairs and discussing their analytical solutions of the equations which are commutative and commutative pairs.

## 3. COMMUTATIVE PAIRS OF SECOND-ORDER DIFFERENTIAL EQUATIONS

We now consider another second-order linear time-varying system $B$ of type $A$ represented by

$$b_2(t)\ddot{y}_B(t) + b_1(t)\dot{y}_B(t) + b_0(t)y_B(t); (t); t \geq 0, \tag{2}$$

with input and output functions $x_B(t)$ and $y_B(t)$, respectively; with the initial conditions $y_B(0)$ and $\dot{y}_B(0)$; and with time-varying coefficients $b_2(t) \not\equiv 0, b_1(t), b_0(t)$.

The set of necessary and sufficient conditions that systems $A$ and $B$ are commutative are

$$\begin{bmatrix} b_2 \\ b_1 \\ b_0 \end{bmatrix} = \begin{bmatrix} a_2 & 0 & 0 \\ a_1 & a_2^{0.5} & 0 \\ a_0 & a_2^{-0.5}(2a_1 - \dot{a}_2)/4 & 1 \end{bmatrix} \begin{bmatrix} c_2 \\ c_1 \\ c_0 \end{bmatrix}, \tag{3a}$$

$$-a_2^{0.5}\frac{d}{dt}\left[a_0 - \frac{1}{16a_2}(4a_1^2 + 3\dot{a}_2^2 - 8a_1\dot{a}_2 + 8\dot{a}_1a_2 - 4a_2\ddot{a}_2)\right]c_1 = 0. \tag{3b}$$

Condition (3a) requires three relations between time-varying coefficients of systems $A$ and $B$, where $c_2, c_1, c_0$ are arbitrary constants. Condition (3b) implies that if $c_1 = 0$, this equation is always satisfied. This special case corresponds to that System $B$ is obtained from System $A$ by constant feed forward path gain $\alpha = 1/c_2$ and negative feedback gain $\beta = c_0$; every System $B$ obtained from $A$ this way is commutative with $A$ without requiring any condition on the coefficients of $A$. This case is considered in the literature [7, 18-21] and being special has not weighted importance for the study in this paper, therefore it is excluded from the general treatment given in the sequel. Otherwise if $c_1 \neq 0$, not all second order systems have commutative pairs; and to have a commutative pair like $B$, the coefficients of $A$ must satisfy Eq. (3b) with the constant $c_1 \neq 0$ in Eq. (3a); this is the case considered in this paper. Note that if $c_2$ in Eq. (3a) is chosen as zero, order of system $B$ reduces to one. When both $c_2$ and $c_1$ are zero, $B$ becomes a scalar (algebraic) system with constant gain $\frac{1}{c_0}, c_0 \neq 0$.

For $c_1 \neq 0$, Eq. (3b) can be written as

$$a_0 - \frac{1}{16a_2}(4a_1^2 + 3\dot{a}_2^2 - 8a_1\dot{a}_2 + 8\dot{a}_1a_2 - 4a_2\ddot{a}_2) = A_0, \tag{3c}$$

where $A_0$ is a constant. The commutativity property of the second-order linear time-varying differential equations listed in Section 2, that is their existence of commutative conjugates other than the above mentioned feed-back structure, are investigated next by using (3c) and the results are listed in Table 2.

**Table 2:** Commutativity property of differential systems described by some famous DE

| Not commutative | Conditionally commutative | Commutative |
|---|---|---|
| 3, 4, 5, 6, 7, 12, 19, 20, 22, 24, 26 | 1, 2, 8, 10, 11, 14, 15, 16, 17, 18, 21, 23, 25, 27, 28, 29, 30 | 9, 13 |

**Example 1:** The first example is for the case of “Not commutative” in Table 1. We assume a system which is modelled by the fourth equation (Bessel differential equation) in the above mentioned list. The coefficients of Eq. (1) are $a_2 = x^2, a_1 = x, a_0 = x^2 - n^2$ for Bessel differential equations. The expression in the bracket in Eq. (3b) should be a constant as in Eq. (3c) for the existence of the commutative pair of a system. But for Bessel equation, writing the coefficients $a_2 = x^2, a_1 = x, a_0 = x^2 - n^2$ and their derivatives in Eq. (3c), this expression becomes $x^2 - n^2$, which is not a constant. Then, a system modelled by Bessel equation listed as the fourth equation in the list does not have any commutative pair.

**Example 2:** The second example is for the case of “Conditionally commutative”. We assume a system $A$ which is modelled by Anger differential equation which is the second equation in the table. The coefficients of Eq. (1) are $a_2 = 1, a_1 = \frac{1}{x}, a_0 = 1 - \frac{v^2}{x^2}$ for Anger differential equation. The expression (3c) which is found as $A_0 = 1 - \frac{v^2}{x^2} + \frac{1}{4x^2}$, should be satisfied for the existence of the commutative pair of $A$. For $v = \pm\frac{1}{2}$, this expression is constant then the equation has commutative pair. Using Eq. (3a) for Anger differential equation where $v = \pm\frac{1}{2}$, the coefficients of its commutative pair are found as follows:

$$b_2 = c_2 a_2 = c_2,$$

$$b_1 = c_2 a_1 + c_1 {a_2}^{0.5} = c_1 + \frac{c_2}{x},$$

$$b_0 = c_2 a_0 + c_1 {a_2}^{-0.5}(2a_1 - \dot{a}_2)\frac{1}{4} + c_0 = c_0 + \frac{c_1}{2x} + c_2\left(1 - \frac{1}{4x^2}\right).$$

Then, commutative pair is written as follows:

$$c_2 y'' + \left(c_1 + \frac{c_2}{x}\right) y' + \left[c_0 + \frac{c_1}{2x} + c_2\left(1 - \frac{1}{4x^2}\right)\right] y = x_2.$$

**Example 3:** The last example is for the case of "Commutative". Let $A$ be a system which is modelled by Chebyshev differential equation (the ninth equation in Table 1). The coefficients of Eq. (1) are $a_2 = 1 - x^2, a_1 = -x, a_0 = n^2$ for Chebyshev differential equation. The expression (3c) yields $A_0 = n^2$ which is always a constant so the system has always a commutative pair. The coefficients of commutative pairs are

$$b_2 = c_2 a_2 = c_2(1 - x^2),$$

$$b_1 = c_2 a_1 + c_1 {a_2}^{0.5} = c_1\sqrt{1 - x^2} - c_2 x,$$

$$b_0 = c_2 a_0 + c_1 {a_2}^{-0.5}(2a_1 - \dot{a}_2)\frac{1}{4} + c_0 = c_0 + c_2 n^2.$$

Then, all the commutative pairs of $A$ described by:

$$c_2(1 - x^2)y'' + \left(c_1\sqrt{1 - x^2} - c_2 x\right)y' + (c_0 + c_2 n^2)y = x_2.$$

In Table 3, commutativity conditions of the conditionally commutative differential systems listed in Table 2 are given and the final form of these equations are presented in the case of the fact that the required condition is used in the original equation stated in Section II. Commutative systems in Table II are repeated in Table III for the sake of completeness. Commutative conjugates of all equations in Table 3 are found by using Eq. (3b) and the results are presented in Table 4.

**Table 3:** Commutativity conditions

| Line # | Name of Equation | Condition for Commutativity | Final Forms of Equations |
|---|---|---|---|
| 1 | Airy DE | $k = 0$ | $y'' = 0$ |
| 2 | Anger DE | $i)\ v = -0.5$<br>$ii) v = 0.5$ | $y'' + \frac{y'}{x} + \left(1 - \frac{1}{4x^2}\right)y = \frac{x + 0.5}{nx^2}$<br>$y'' + \frac{y'}{x} + \left(1 - \frac{1}{4x^2}\right)y = \frac{x - 0.5}{nx^2}$ |
| 3 | Baer DE | $p = 0$ | $(x-d_1)(x-d_2)y'' + \frac{1}{2}[2x-(d_1+d_2)]y' - q^2 y = 0$ |
| 4 | Bessel DE-wave | $a = b = 0$ | $x^2 y'' + xy' - c^2 y = 0$ |
| 5 | Chebyshev DE | no condition | $(1-x^2)y'' - xy' + n^2 y = 0, where \|x\| < 1$ |
| 6 | Eckart DE | $i)\ \alpha = \beta = \gamma = 0$<br>$ii) \delta = 0$ | $y'' = 0$<br>$y'' + \left(\frac{\alpha\eta}{2} + \frac{\beta\eta}{4} + \gamma\right) y = 0$ |
| 7 | Ellipsoidal wave DE | $i)\ k = 0$<br>$ii) b = q = 0$ | $y'' - ay = 0$ |
| 8 | Euler DE | no condition | $x^2 y'' + \alpha xy' + \beta y = s(x)$ |
| 9 | Gegenbauer DE | $i)\ \mu = -0.5$<br>$ii) \mu = 0.5$ | $(1-x^2)y'' - xy' + (v+0.5)^2 y = 0$<br>$(1-x^2)y'' - 3xy' + (v-0.5)(v+1.5)y = 0$ |
| 10 | Hill's DE | $\theta_n = 0\ for\ n > 0$ | $y'' + \theta_0 y = 0$ |
| 11 | Hypergeometric DE | $i)\ c = 0.5, a + b = 0$<br>$ii)\ c = 0.5, a + b = 1$<br>$iii) c = 1.5, a + b = 2$<br>$iv) c = 1.5, a + b = 1$ | $x(1-x)y'' + (0.5-x)y' + a^2 y = 0$<br>$x(1-x)y'' + (0.5-2x)y' - a(1-a)y = 0$<br>$x(1-x)y'' + (1.5-3x)y' - a(2-a)y = 0$<br>$x(1-x)y'' + (1.5-2x)y' - a(1-a)y = 0$ |
| 12 | Jacobi DE-first | $i)\ \alpha = \beta = -0.5$<br>$ii)\ \alpha = \beta = 0.5$<br>or<br>$iii) \alpha = 0.5, \beta = -0.5$<br>$iv) \alpha = -0.5, \beta = 0.5$ | $(1-x^2)y'' - xy' + n^2 y = 0$<br>$(1-x^2)y'' - 3xy' + n(n+2)y = 0$<br><br>$(1-x^2)y'' - (1+2x)y' + n(n+1)y = 0$<br>$(1-x^2)y'' + (1-2x)y' + n(n+1)y = 0$ |
| 13 | Jacobi DE-second | $i)\ \alpha = 0, \beta = 0.5$<br>$ii)\ \alpha = 1, \beta = 0.5$<br>$iii) \alpha = 1, \beta = 1.5$<br>$iii) \alpha = 2, \beta = 1.5$ | $x(1-x)y'' + (0.5-x)y' + n^2 y = 0$<br>$x(1-x)y'' + (0.5-2x)y' + n(n+1)y = 0$<br>$x(1-x)y'' + (1.5-2x)y' + n(n+1)y = 0$<br>$x(1-x)y'' + (1.5-2x)y' + n(n+2)y = 0$ |
| 14 | Morse-Rosen DE | $\alpha = 0$ | $y'' + \gamma y = 0$ |
| 15 | Parabolic Cylinder DE | $a = b = 0$ | $y'' + cy = 0$ |
| 16 | Richardson’s DE | $\lambda = 0$ | $y'' + \mu y = 0$ |
| 17 | Symmetric top DE | $M = 0, K = 0.5$ | $y'' - (\delta + 0.5)y = 0$ |
| 18 | Titchmarsh’s DE | $n = 0$ | $y'' + \lambda y = 0$ |
| 19 | Weber DE-first | $b = 0$ | $y'' + a^2 y = 0$ |
| 20 | Weber DE-second | $i)\ v = -0.5$<br><br>$ii) v = 0.5$ | $y'' + \frac{1}{x}y' + \left(1 - \frac{1}{4x^2}\right)y = -\frac{x - 0.5}{\pi x^2}$<br><br>$y'' + \frac{1}{x}y' + \left(1 - \frac{1}{4x^2}\right)y = -\frac{x + 0.5}{\pi x^2}$ |

**Table 4:** Commutative conjugates of differential equations in Table 3

| **Name of Equation** | **Commutativity Conjugates** |
|---|---|
| Airy DE | $c_2 y'' + c_1 y' + c_0 y = x_2$ |
| Anger DE | $c_2 y'' + \left(\frac{c_2}{x} + c_1\right) y' + \left[c_2\left(1 - \frac{1}{4x^2}\right) + \frac{c_1}{2x} + c_0\right] y = x_2$ |
| Baer DE | $c_2(x - d_1)(x - d_2)y'' + \left\{c_2[x - 0.5(d_1 + d_2)] + c_1\sqrt{(x - d_1)(x - d_2)}\right\} y'$<br>$+ (-c_2 q^2 + c_0)y = x_2$ |
| Bessel DE-wave | $c_2 x^2 y'' + (c_2 + c_1)xy' + (-c_2 c^2 + c_0)y = x_2$ |
| Chebyshev DE | $(1 - x^2)y'' - xy' + n^2 y = 0, where\ \lvert x \rvert < 1$ |
| Eckart DE | $i)\ c_2 y'' + c_1 y' + c_0 y = x_2$<br>$ii)\ c_2 y'' + c_1 y' + \left[c_2\left(\frac{\alpha\eta}{2} + \frac{\beta\eta}{4} + \gamma\right) + c_0\right] y = x_2$ |
| Ellipsoidal wave DE | $c_2 y'' + c_1 y' + (-c_2 a + c_0)y = x_2$ |
| Euler DE | $c_2 x^2 y'' + (c_2 \alpha + c_1)xy' + \left[c_2 \beta + c_1(\alpha - 1)\frac{1}{2} + c_0\right] y = x_2$ |
| Gegenbauer DE | $i)\ c_2(1 - x^2)y'' + \left[c_2 x + c_1\sqrt{1 - x^2}\right] y' + \left[c_2(v + 0.5)^2 + \frac{c_1 x}{\sqrt{1 - x^2}} + c_0\right] y$<br>$= x_2$ |
| | $ii)\ c_2(1 - x^2)y'' + \left[-3c_2 x + c_1\sqrt{1 - x^2}\right] y'$<br>$+ \left[c_2(v - 0.5)(v + 1.5) + \frac{c_1 x}{\sqrt{1 - x^2}} + c_0\right] y = x_2$ |
| Hill’s DE | $c_2 y'' + c_1 y' + (c_2 \theta_0 + c_0)y = x_2$ |
| Hypergeometric DE | $i) c_2(x-x^2)y'' + \left[c_2(0.5-x) + c_1\sqrt{x-x^2}\right]y' + [c_2a^2 + c_0]y = x_2$ |
| | $ii) c_2(x-x^2)y'' + \left[c_2(0.5-2x) + c_1\sqrt{x-x^2}\right]y' + \left[c_2(a^2-a) - c_1\frac{x}{2\sqrt{x-x^2}} + c_0\right]y = x_2$ |
| | $iii) c_2(x-x^2)y'' + \left[c_2(1.5-3x) + c_1\sqrt{x-x^2}\right]y' + \left[c_2(a^2-2a) + c_1\frac{2x-1}{2\sqrt{x-x^2}} + c_0\right]y = x_2$ |
| | $iv) c_2(x-x^2)y'' + \left[c_2(1.5-2x) + c_1\sqrt{x-x^2}\right]y' + \left[c_2(a^2-a) + c_1\frac{1-x}{2\sqrt{x-x^2}} + c_0\right]y = x_2$ |
| Jacobi DE-first | $i) c_2(1-x^2)y'' + \left[-c_2x + c_1\sqrt{1-x^2}\right]y' + (c_2n^2 + c_0)y = x_2$ |
| | $ii) c_2(1-x^2)y'' + \left[-3c_2x + c_1\sqrt{1-x^2}\right]y' + \left[c_2(n^2+n) - c_1\frac{x}{\sqrt{1-x^2}} + c_0\right]y = x_2$ |
| | $iii) c_2(1-x^2)y'' + \left[c_2(1+2x) + c_1\sqrt{1-x^2}\right]y' + \left[c_2(n^2+n) - c_1\frac{1+x}{2\sqrt{1-x^2}} + c_0\right]y = x_2$ |
| | $iv) c_2(1-x^2)y'' + \left[c_2(1-2x) + c_1\sqrt{1-x^2}\right]y' + \left[c_2(n^2+n) + c_1\frac{1-x}{2\sqrt{1-x^2}} + c_0\right]y = x_2$ |
| Jacobi DE-second | $i) c_2(x-x^2)y'' + \left[c_2(0.5-x) + c_1\sqrt{x-x^2}\right]y' + (c_2n^2 + c_0)y = x_2$ |
| | $ii) c_2(x-x^2)y'' + \left[c_2(0.5-2x) + c_1\sqrt{x-x^2}\right]y' + \left[c_2(n^2+n) - c_1\frac{x}{2\sqrt{x-x^2}} + c_0\right]y = x_2$ |
| | $iii) c_2(x-x^2)y'' + \left[c_2(1.5-2x) + c_1\sqrt{x-x^2}\right]y' + \left[c_2(n^2+n) + c_1\frac{1-x}{2\sqrt{x-x^2}} + c_0\right]y = x_2$ |
| | $iv) c_2(x-x^2)y'' + \left[c_2(1.5-2x) + c_1\sqrt{x-x^2}\right]y' + \left[c_2(n^2+n) + c_1\frac{1-x}{2\sqrt{1-x^2}} + c_0\right]y = x_2$ |
| Morse-Rosen DE | $c_2y'' + c_1y' + (c_2\gamma + c_0)y = x_2$ |
| Parabolic Cylinder DE | $c_2y'' + c_1y' + (c_2c + c_0)y = x_2$ |
| Richardson's DE | $c_2y'' + c_1y' + (c_2\mu + c_0)y = x_2$ |
| Symmetric top DE | $c_2y'' + c_1y' + [-c_2(\delta + 0.5)\mu + c_0]y = x_2$ |
| Titchmarsh's DE | $c_2y'' + c_1y' + (c_2\lambda + c_0)y = x_2$ |
| Weber DE-first | $c_2y'' + c_1y' + (c_2a^2 + c_0)y = x_2$ |
| Weber DE-second | $c_2y'' + \left[\frac{c_2}{x} + c_1\right]y' + \left[c_2\left(1 - \frac{1}{4x^2}\right) + \frac{c_1}{2x} + c_0\right]y = x_2$ |

## 4. SOLVABILITY

Ten of final forms of the equations in the third column of Table 3 (Lines 1, 6, 7, 10, 14, 15, 16, 17, 18, 19) are in the form of Airy differential equation. The others are in the form of inhomogeneous Bessel DE (Lines 2, 20), Baer DE (Line 3), Euler DE (Lines 4, 8), Jacobi DE–first (Lines 9, 12) and Hypergeometric DE (Lines 11, 13). Note that Gegenbauer DE (Line 9) and its more general case Jacobi DE-first (Line 12) can be transformed to Hypergeometric DE. Generally, final forms of equations in Table 3 are written in the form $(ax^2 + bx + c)y'' + (dx + e)y' + (x^2 - p^2)y = f(x)$ and they can also also be solved analytically by using special functions and polynomials generally famed by the person who defined the function. For a more detail of analytical solutions of these equations, we operate the reader to the first subsection entitled "Linear Equations" of the second Chapter entitled "Second-order Differential Equations" of the reference [22] which contains a great deal of exact solutions methods for ordinary differential equations. Particular solutions of (homogeneous and nonhomogeneous) some differential equations are given and the way for obtaining the general solution of the nonhomogeneous equations after finding another linearly independent particular solution by using the given particular solution is explained with detail.

On the other hand, commutative pairs shown in Table 4 are found by using Eq. (3a). The solvability of commutative pairs listed in Table 4 are discussed in the sequel. Commutative conjugates of Airy DE, Eckart DE, Ellipsoidal wave DE, Hill's DE, Morse-Rosen, DE, Parabolic Cylinder DE, Richordsen's DE, Symmetric top DE, Titchmarsh's DE and Weber DE-first are second order linear differential equations with constant coefficient. They can be solved analytically by using explicit methods. The forms of commutative conjugates of Bessel DE-wave and Euler DE are the same with each other. Moreover, the forms of these conjugates are also same with the final forms of these equations and those can be solved analytically. The forms of commutative conjugates of Anger DE and Weber DE are also same and those can also be solved analytically. The details of how all these equations can be solved are found from the first section of the second chapter of the book [22]. As shown in Table 4, some expressions containing square roots of functions depending on independent variable exit as the coefficients of $y$ and $y'$ in commutative conjugates of Baer DE, Chebyshev DE, Gegenbauer DE, Hypergeometric DE and Jacobi Des. According to the authors' search, there is no specific method for solving the commutative conjugates of these equations analytically.

## 5. USE OF COMMUTATIVITY FOR CRYPTOLOGY IN SECRET COMMUNICATION

Some benefits of commutativity of linear time-varying systems have already been appeared in the literature; for example designing systems less sensitive to parameter values [21], reducing noise interference and disturbance [22] improving robustness [6].

This article focuses attention a new encrypting method of obscuring the information transmitted through any communication channel by disguising it between transmitter and receiver. More precisely consider a communication system as shown in Fig. 2.

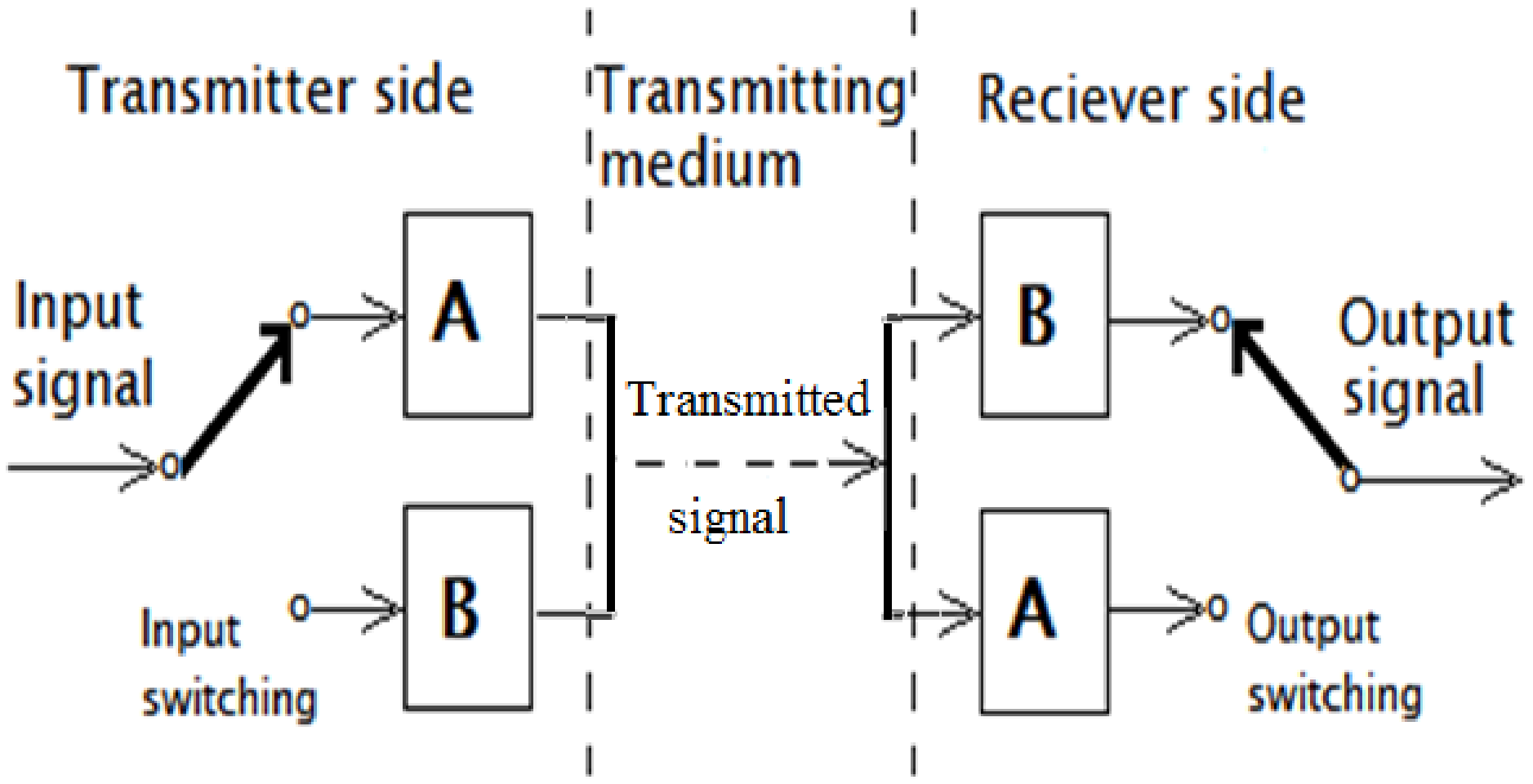

**Figure 2:** Transmitting a secret signal in different forms through a transmission channel.

In the figure, $A$ and $B$ represent commutative linear time-varying systems so that both channels AB and BA produce the same output signal for any applied input signal. But the transferred signal from transmitter to receiver proceed in completely different shapes through the transmitting medium. Hence, this generates somewhat prevention against the infiltrators to stealing the secret information during transmission. The concept can be extended to more complex structures by using higher number of switching greater than 1. For example, with two identical subsystems $A$ and two identical subsystems $B$ (commutative with $A$) 4 communication passages of the input signal can be achieved through transmitting medium to obtain the same output signal. In fact, the structures $A \rightarrow ABB$, $AA \rightarrow BB$, $AAB \rightarrow B$, $AB \rightarrow AB$ where the arrow " $\rightarrow$ " separates the subsystems appearing in the transmitter and receiver sides. All these structures transfer any input signal to the same output signal which is transmitted in different shapes in transmitting medium by all of four structures. The concept can be extended for more complicated cases by using more than two different commutative pairs.

To see how any input signal is transmitted to the same output signal in different forms of the transmitting medium, consider the communication structure in Fig. 3. With the linear time-varying subsystems A and B described by

$$A\colon\ \ddot{y}_A + (2 + 2sinw_0t)\dot{y}_A + \left(5 - \frac{1}{2}cos2w_0t + 2sinw_0t + w_0cosw_0t\right)y_A = x_A, \quad (5.1)$$

$$B\colon \frac{1}{2}\,\ddot{y}_B + \left(\frac{3}{4} + sinw_0t\right)\dot{y}_B$$

$$+\left(\frac{409}{32} - \frac{1}{4}cos2w_0t + \frac{3}{4}sinw_0t + \frac{1}{2}w_0cosw_0t\right)y_B = x_B, \quad (5.2)$$

where $x_i$ and $y_i$ represent the input and output, respectively, of the subsystems $i = A, B$: (Double) dot on the top indicates (second) time derivative.

It is straight forward to show that A and B are commutative since the time-varying coefficients of $B$ can be obtained from those of A by the relation (3a) in [22]

$$\begin{bmatrix} b_2(t) \\ b_1(t) \\ b_0(t) \end{bmatrix} = \begin{bmatrix} a_2(t) & 0 & 0 \\ a_1(t) & {a_2}^{0,5}(t) & 0 \\ a_0(t) & f_A(t) & 1 \end{bmatrix} \begin{bmatrix} k_2 \\ k_1 \\ k_0 \end{bmatrix},$$

where $k_2 = \frac{1}{2}, k_1 = -\frac{1}{4}, k_0 = \frac{4213}{400}$ and

$$f_A = \frac{2a_1 - \dot{a}_2}{4\sqrt{a_2}} = 1 + sinw_0t.$$

Since $k_1 \neq 0$, the second one of the sufficient conditions of commutativity

$$A_0(t) = a_0 - f_A^2 - \sqrt{a_2} f_A(t) = 3.5$$

is satisfied since $A_0(t)$ in Eq. (3c) is constant (See Eq. (2.b). in [22]). It is easy to show that when the average values of coefficients are considered both systems are asymptotically stable with eigenvalues

$$A_{1,2} = -1 \pm j2,$$

$$B_{1,2} = -\frac{3}{4} \pm j5,$$

This implies though not guaranties, the high possibility of stability of actual time-varying subsystems $A$ and $B$ defined by Eqs. (5.1) and (5.2), respectively [23]; in fact, simulation results show that both systems are asymptotically stable.

To observe that both of the switching alternatives $A \to B$ and $B \to A$ shown in Figure 2 yield the same output at the receiver side, an input signal ($30sin1.2\pi t$ + a saw-tooth of period 3.3s and increasing from $-30$ to $+30$) is applied on the transmitter side. As observed in Figure 3 the transmissions $A \to B$ and $B \to A$ yield the same output signal (see output signal *10). In spite of the same input-output pairs for switchings $AB$ and $BA,$ the travelled signals processed through transmission medium (see Transmitted signal A-B, Transmitted signal B-A) are quite different.

To verify that the discussions are independent of the input signal applied, the simulations are repeated with a pulse train of amplitude 30, period 5 and with a pulse with of 10 %. The input signal and the same output of both transmission switching paths $A \to B$ and $B \to A$ are shown in Figure 4 (see Input signal/10, Output signal*10, respectively). It is also seen in this figure that the signals proceeded through the transmission medium, namely Transmitted s. A-B and Transmitted signal B-A, are quite different. Hence the same output signal is received by channels $AB$ and $BA$ for the same input signals irrespective of the shape of the input signal whilst different signals are transmitted through the transmission medium.

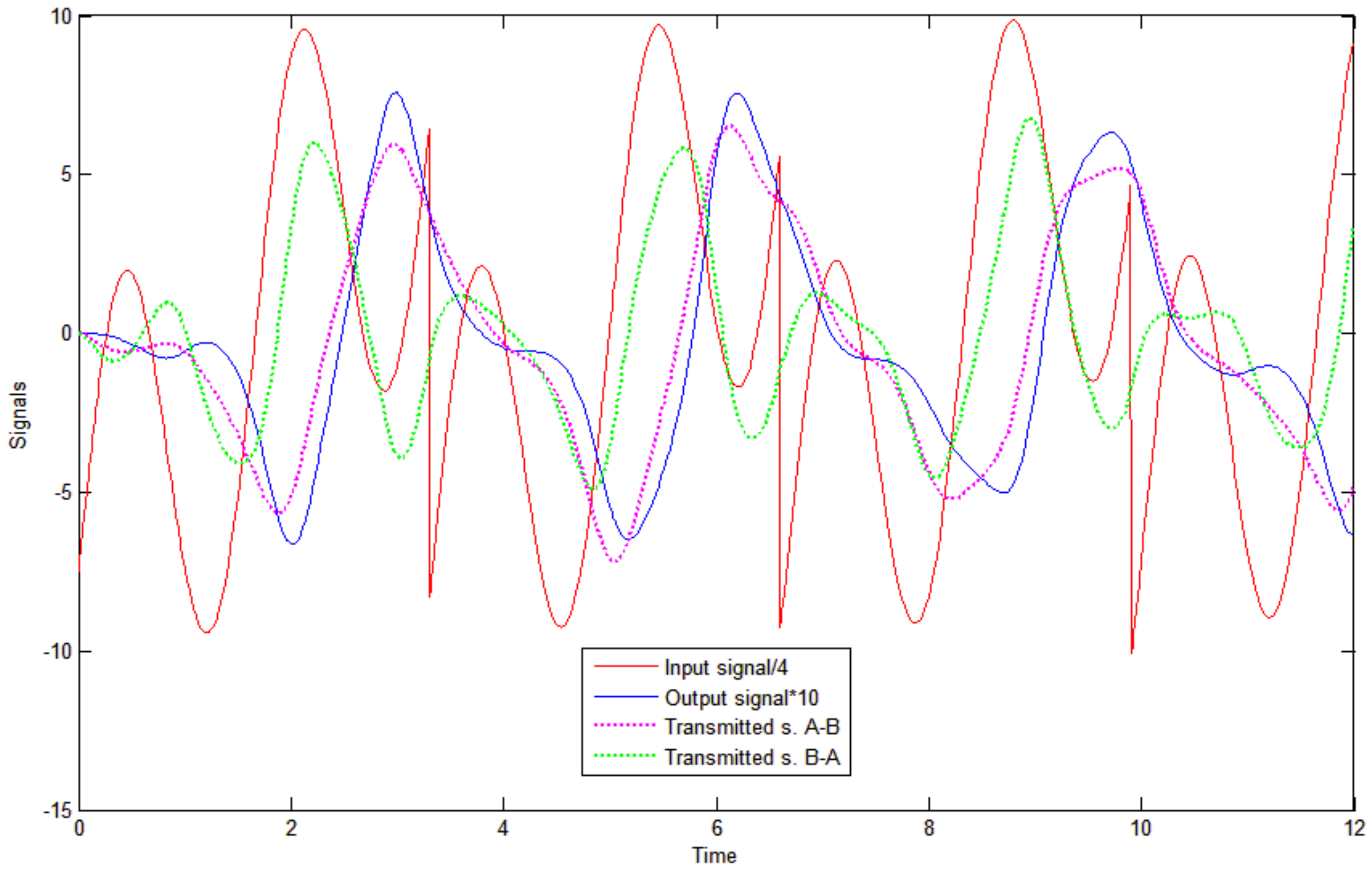

**Figure 3:** Input, output and transmitted signals in a communication system.

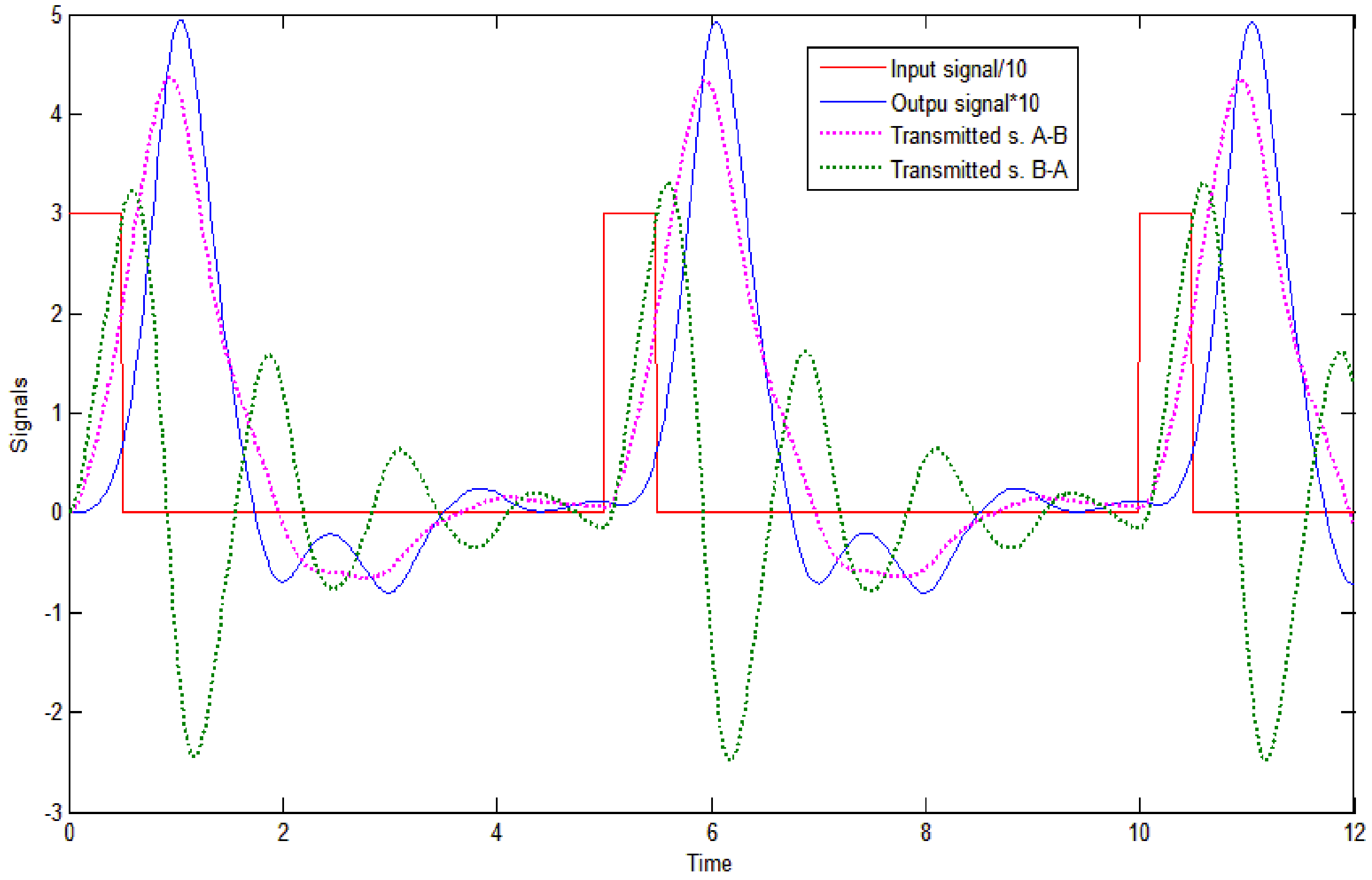

**Figure 4:** Input, output and transmitted signals in a communication system for a different input.

## 6. DISCUSSION AND CONCLUSION

In this study, second-order linear differential equations with variable coefficients are searched in the literature and 30 second-order linear differential equations are presented. The existences of their commutative pairs are investigated. It is shown that some of them have always commutative conjugates, some of them have commutative conjugates under certain conditions found in the paper and some of them do not have any commutative conjugates at all. Commutative conjugates of differential equations whose commutativity pairs exit are constructed. Moreover, their solvability are discussed.

There are much more than 30 second-order linear differential equations exiting in the literature records but it is impossible to investigate all of them in a paper. Investigation of commutativity of these equations and solvability of their commutative pairs, if they exit, can be studied as future work.

Among commutative conjugates of differential systems appearing in Table 4, some of them do not have explicit solutions and the search for explicit solutions constitutes another future work.

A new possible use of commutativity is illustrated showing how transmitted signals transmission medium can be changed without affecting the received signal, a fact which can be used for preventing the actual signal to be revealed easily by unauthorized persons.

**Acknowledgment:** This work is supported by the Scientific and Technological Research Council of Turkey (TUBITAK) under the project no. 115E952.